\documentclass[prc,twocolumn,showpacs,preprintnumbers,amsmath,amssymb,superscriptaddress,floatfix,nofootinbib]{revtex4}
\usepackage{graphicx} 
\usepackage{hyperref}
\hypersetup{hypertex=true,colorlinks=true,linkcolor=blue,anchorcolor=blue,citecolor=blue}

\begin{document}

\title{Study on charmonium(-like) mesons within a diabatic approach
}
\author{Zi-Zhao Zhang}\author{Rong Li}\author{Bo-Chao Liu}
\email{liubc@xjtu.edu.cn}
\affiliation{MOE Key Laboratory for Nonequilibrium Synthesis and Modulation of Condensed Matter, School
of Physics, Xi’an Jiaotong University, Xi ’an 710049, China.}
\begin{abstract}
 In this work, we study the charmonium(-like) spectrum below 4.1 GeV using the diabatic approach, which offers a unified description of conventional and unconventional heavy meson states. Compared to previous studies, we consider a more realistic $c\bar c$ potential with including the spin-dependent interactions, which allows us to obtain more states and get more insights on the charmonium spectrum. Based on our calculation, we obtain the masses of the charmonium spectrum which align with the experimental data well. We also present the probabilities of finding various components, i.e. $c\bar c$ or meson-meson pair, in those states. Our results support the arguments that the $\chi_{c1}(3872)$, $\psi(4040)$ and $\chi_{c2}(3930)$ have significant molecular components. In addition, our calculations show that the $\chi_{c0}(3860)$ and $\psi(3770)$ can be looked as the candidates for the charmonium states $\chi_{c0}(2P)$ and $\psi(1D)$, respectively.

\end{abstract}
\maketitle
\section{Introduction}
 Studies on charmonium(-like) states play a significant role in understanding how quarks constitute hadrons, which is closely related to the non-perturbative behavior of strong interactions. Since the discovery of the first charmonium meson $J/\psi$\cite{PhysRevLett.33.1404,SLAC-SP-017:1974ind} in 1974, a series of charmonium mesons, such as $\psi(3686)$ \cite{osti_4239059},$\chi_{c0}(1P)$ \cite{PhysRevLett.38.1324},$\chi_{c2}(1P)$ \cite{PhysRevLett.37.1596} and others\cite{PhysRevLett.35.1323,PhysRevLett.45.1150} were successively discovered in experiments. To understand the properties of these states, various potential models were proposed to describe the interactions between quarks \cite{PhysRevD.17.3090,VOLOSHIN2008455}. Among them, the Cornell potential \cite{Eichten:1974af} is widely adopted to predict and interpret the charmonium spectrum. While the Cornell potential is rather successful in describing the low lying charmonium states,
 its predictions for many highly excited states are still controversial and not well identified in experiments. At the same time, in the past two decades, a series of charmonium-like states called XYZ states have been discovered experimentally\cite{Brambilla:2019esw}, whose masses differ significantly from the predictions of the potential model \cite{barnes2005higher}. Some of them clearly do not fit with the conventional $q\bar q$ picture of meson state(see Refs.\cite{RevModPhys.90.015004,Ji:2022vdj,Hanhart:2019isz,Chen:2022asf} for recent reviews).

For the newly discovered charmoinum-like states, a special feature is that the masses of many states lie very close to some meson pair thresholds, which have caused the arguments on their molecular natures. In general, for higher excited states the closeness to the open charm thresholds may significantly affect the charmonium spectrum through the virtual meson loop. To describe these states, it is then necessary to take into account such coupled-channel effects. Along this line, various approaches were developed\cite{PhysRevD.80.014012,PhysRevD.72.034010,PhysRevD.17.3090,PhysRevD.76.077502,ono1984continuum}. One solution is to introduce the adiabatic approximation.

 The adiabatic approximation, also known as the Born-Oppenheimer (B-O) approximation, was initially developed to describe molecular systems \cite{born1985quantentheorie} and has since found extensive applications in molecular and atomic physics as well as meson states in QCD \cite{bali2001qcd, castella2022heavy}. While, the B-O approximation relies on two key assumptions: the adiabatic approximation and the single-channel approximation\cite{PhysRevD.102.074002,Lebed:2022vks,Lebed:2024rsi}. As mentioned above, for charmonium states being close to the open flavor meson-meson threshold, the mixing between the $Q\bar Q$ and meson-meson configurations cannot be ignored, which challenges the validity of the single-channel approximation assumed in the B-O approximation. In this case, the system should be described by a set of coupled-channel Schrödinger equations containing non-adiabatic coupling terms (NACTs)\cite{beyondBorn-Oppenheimer}, which in practice are difficult to solve\cite{PhysRevD.102.074002}. To overcome the problems, it is required to generalize the B-O approximation and adopt the so-called diabatic formalism.

 The diabatic formalism has many advantages. Firstly, in this approach the dynamics is described by a diabatic potential matrix, which in principle can be calculated directly in lattice QCD and provides a non-perturbative method to take into account the mixing of different configurations. Therefore, it offers a way to study the heavy quarkonium(-like) states based on the first principles or models inspired by the Lattice QCD results. In practice, once the diabatic potential matrix is determined, one may use numerical methods to solve the coupled-channel Schrödinger equation to obtain the solution of the system.
 In addition, by choosing appropriate expansion basis, it is possible to make the wave function components correspond to a pure $Q\overline{Q}$ state or the meson-meson state. In this way, the physical meaning of each channel is intuitive, which makes it more convenient to analyze the results.

 In previous studies\cite{PhysRevD.102.074002,Bruschini_2021}, the authors apply the diabatic approach to study quarkonium spectrum with masses below 4.1 GeV. With using Cornell potential to describe the $c\overline{c}$ interaction, they calculated the charmonium meson spectrums of $0^{++}$, $1^{++}$, $2^{++}$ and $1^{--}$ states. However, in their works the spin-dependent interaction between heavy quarks was not considered. The absence of spin dependent terms prevented the authors from exploring the fine and hyperfine structures in the charmonium spectrum, which is however important for a complete and accurate understanding on the charmonium spectrum.

 In this work, we extend the works in Refs.\cite{PhysRevD.102.074002,Bruschini_2021} by further considering the spin-dependent interactions between the $c\overline{c}$, which cause the fine and hyperfine splittings of the chamonium mesons. By solving the coupled-channel Schrödinger equations, we get the
 masses of charmonium spectrum below 4.1 GeV and the probabilities of finding $c\bar c$ or various meson pair components in those states. Comparing to the results without considering the spin-dependent interactions, we can now obtain more states to compare with experimental data. Furthermore, it also leads to a better description of lower lying states in the charmonium spectrum.
 Since this formalism provides a unified description of both conventional and unconventional heavy meson states, we can calculate and describe the states like the  $\chi_{c0}(3860)$, for which there are still disputes on their natures\cite{Petrov:2005tp,Yu:2017bsj}. Through analyzing the probability of different configurations, we can explore whether a state is a conventional state or not.  Hence, our results may provide new insights on some controversial charmonium(-like) states.

 This paper is organized as follows. In Section II we review  the formalism of the diabatic approach which is developed from the B-O approximation. In Section III, we construct the potential matrix of the system with considering spin-dependent terms. In Section IV, we apply the diabatic approach to study the charmonium spectrum and discuss the results. Finally, we summarize our conclusions in Section V.

\section{DIABATIC FORMALISM}
 The Hamiltonian for a system of heavy quarkonium state can be written as\cite{PhysRevD.102.074002} \begin{equation}
H=\frac{\boldsymbol{p}^2}{2\mu_{Q\overline{Q}}}+\frac{\boldsymbol{P}^2}{2(m_Q+m_{\bar Q})}+H^\mathrm{light}\label{1}
\end{equation} with $\mu_{Q\overline{Q}}$ being the reduced mass of the $Q\bar Q$ system, $\boldsymbol{p}(\boldsymbol{P})$ the $Q\bar Q$ relative(total) three momentum. Here we use $H^\mathrm{light}$ to represent the Hamiltonians describing the energy of light fields like light quarks and gluons and their interaction with the $Q\bar Q$. The Schrödinger equation of the system in the center-of-mass frame of $Q\bar Q$ can be written as
 \begin{equation}
(\frac{\boldsymbol{p}^2}{2\mu_{Q\overline{Q}}}+H^\mathrm{light}-E)|\psi\rangle=0.\label{se}
\end{equation}
 In the heavy quarkonium system, the mass of the heavy quark is significantly larger than the energy scale of the light fields, which justifies the neglect of the kinetic energy of the heavy field, i.e. using the static limit, when considering the dynamics of the light fields. As a result, the separation of the heavy quarks $\boldsymbol{r}$ can be regarded as a c-number parameter rather than a dynamical operator. Hence, the Hamiltonian $H^{\mathbf{light}}$ becomes an operator, $H_{static}^{\mathrm{light}}(\boldsymbol{r})$, solely pertaining to the light field, and now the $\boldsymbol{r}$ is a parameter denoting the separation of the heavy quarks. So, in the static limit, the dynamics of light fields in the $Q\bar Q$ rest frame is described by \begin{equation}
  (H_{\mathrm{static}}^{\mathrm{light}}(\boldsymbol{r})-V_i(\boldsymbol{r}))|\zeta_i(\boldsymbol{r})\rangle=0,\label{light}
  \end{equation}
  where $V_i(\boldsymbol{r})$ is the  eigenvalue of the reduced Hamiltonian $H_{\mathrm{static}}^{\mathrm{light}}(\boldsymbol{r})$.

In order to solve Eq.\eqref{se}, we apply the diabatic expansion to the eigenstates $|\psi\rangle$ as in Ref.\cite{PhysRevD.102.074002}(see also Refs.\cite{ Lebed:2022vks,Lebed:2024rsi}). In brief, we use the  $|\zeta_i(\boldsymbol{r_0})\rangle$, which is the i-th eigenstate of $H^{light}_{static}(\boldsymbol{r_0})$ with $\boldsymbol{r_0}$ being a free parameter and taking some fixed value, to expand the state $|\psi\rangle$ and get
\begin{equation}
|\psi\rangle=\sum_i\int\mathrm{d}\boldsymbol{r}^{\prime}\tilde{\psi}_i(\boldsymbol{r}^{\prime},\boldsymbol{r}_0)|\boldsymbol{r}^{\prime}\rangle|\zeta_i(\boldsymbol{r}_0)\rangle, \label{12}
\end{equation}
with the $\tilde{\psi}_i(\boldsymbol{r}^{\prime},\boldsymbol{r}_0)$ being the wave function corresponding to the i-th light field state and the $\boldsymbol{r}^{\prime}$ standing for the separation between the heavy quarks. Substituting Eq.\eqref{12} into Eq.\eqref{se} and multiplying  $\langle\boldsymbol{r}|$ and $\langle\zeta_j(\boldsymbol{r_0})|$ on the left side of the equation give
\begin{equation}
\sum_i\left(-\frac{\hbar^2}{2\mu}\delta_{ji}\nabla^2+V_{ji}(\boldsymbol{r},\boldsymbol{r}_0)-E\delta_{ji}\right)\tilde{\psi}_i(\boldsymbol{r},\boldsymbol{r}_0)=0,\label{13}
\end{equation}
with the diabatic potential matrix being defined as
\begin{equation}
V_{ji}(\boldsymbol{r},\boldsymbol{r}_0)\equiv\langle\zeta_j(\boldsymbol{r}_0)|H_{static}^\mathrm{light}(\boldsymbol{r})|\zeta_i(\boldsymbol{r}_0)\rangle. \label{14}
\end{equation}

 The introduction of the diabatic potential matrix is a crucial step in the diabatic approach. Since the $r_0$ is a free parameter, it is convenient to choose an appropriate $\boldsymbol{r_0}$ so that the light field configuration, i.e. each state $|\zeta_i(\boldsymbol{r}_0)\rangle$, corresponds to a pure $Q\overline{Q}$ state or the meson-meson state. In this way, the physical meaning of each state is clear, which makes the analysis more intuitive.
 For later convenience, we relabel $Q\overline{Q}$ state as the 0 state and use $n$ with its value starting from 1 to label the n-th meson-meson pair state considered in this work. With these substitutions, we then have
\begin{equation}
|\zeta_0(\boldsymbol{{r}_0})\rangle\to|\zeta_{Q\overline{Q}}\rangle,\quad|\zeta_n(\boldsymbol{{r}_0})\rangle\to|\zeta_{M{\overline M}_{n}}\rangle,\label{15}
\end{equation}
 to denote various states. And the corresponding wave functions are denoted as
 \begin{equation}
    \tilde{\psi}_0(\boldsymbol{{r},{r}_0})\to{\psi}_{{Q}\overline{Q}}(\boldsymbol{r}),\quad\tilde{{\psi}}_n(\boldsymbol{{r},{r}_0})\to{\psi}^{M\overline{{M}}}_{n}(\boldsymbol{r}).\label{16}
 \end{equation} \par
The matrix element of the $Q\bar Q$ interaction is
 \begin{equation}
    V_{00}(\boldsymbol{r},\boldsymbol{r}_0)\to V_{Q\overline{Q}}(\boldsymbol{r})=\langle\zeta_{Q\overline{Q}}|H_{\mathrm{static}}^\mathrm{light}(\boldsymbol{r})|\zeta_{Q\overline{Q}}\rangle. \label{17}
 \end{equation}
 The matrix elements describing the interactions of the meson-meson pairs are
 \begin{equation}
    V_{ij}(\boldsymbol{r},\boldsymbol{r}_0)\to V^{M\overline{M}}_{ij}(\boldsymbol{r})=\langle\zeta_{M\overline{M}_i}|H_{\mathrm{static}}^{\mathrm{light}}(\boldsymbol{r})|\zeta_{M\overline{M}_j}\rangle,\label{9}
 \end{equation}
 with $1\leq i,j \leq N$ and N being the total number of $M\bar M$ states considered in this work.
 The matrix element of the mixing potential is denoted as
 \begin{equation}
    V_{0j}(\boldsymbol{r},\boldsymbol{r}_0)\to V_{mix}(\boldsymbol{r})=\langle\zeta_{Q\overline{Q}}|H_{\mathrm{static}}^{\mathrm{light}}(\boldsymbol{r})|\zeta_{M\overline{M}_j}\rangle, \label{19}
 \end{equation}
with $1\leq j \leq N$ and  $V_{0j}(\boldsymbol{r},\boldsymbol{r}_0)= V_{j0}(\boldsymbol{r},\boldsymbol{r}_0)$.

 With these notations, Eq.\eqref{13} can be rewritten in a matrix form,
 \begin{equation}
 (\mathbf{K}+\mathbf{V}(r))\mathbf{\Psi}(r)=E\mathbf{\Psi}(r),\label{33}
 \end{equation}
 where $\mathbf{K}$ is the matrix composed of kinetic energy terms, $\mathbf{V}(r)$ is the potential matrix and $\mathbf{\Psi}(r)$ is a column vector of the wave functions. The explicit form of $\mathbf{K}$ is
\begin{equation}
     \mathbf{K}=
     \begin{bmatrix}
         -\frac{\hbar^{2}}{2\mu_{c\bar c}}\nabla^{2}&\\
         &-\frac{\hbar^2}{2\mu_{M\overline{M}}^{(1)}}\nabla^2\\
         & &\ddots\\
         & & &-\frac{\hbar^2}{2\mu_{M\overline{M}}^{(N)}}\nabla^2
     \end{bmatrix}
 \end{equation}
 where $\mu_{c\bar c}$ is the reduced mass of the $c\overline{c}$, $\mu_{M\overline{M}}^{(i)}$($1\le i\le N$) is the reduced mass of i-th meson-meson pair.

 Analogous to lattice-QCD studies \cite{Bulava_2019}, we neglect
interactions between different meson-meson components, which yields $V_{ij}(\boldsymbol{r},\boldsymbol{r}_0)=0$ for $i\ne j$. The explicit form of $\mathbf{V}(r)$ can be given as
 \begin{equation}
     \mathbf{V}(r)=
     \begin{bmatrix}
         V_{c\overline{c}}(r)&V^{mix}_1(r)&\cdots&V^{mix}_N(r)\\
         V^{mix}_1(r)& V^{M\overline{M}}_{11}&\\
         \vdots& & \ddots\\
         V^{mix}_N(r)& & & V^{M\overline{M}}_{NN}
     \end{bmatrix}.
 \end{equation}
 And the $\mathbf{\Psi}(r)$ is defined as
 \begin{equation}
     \mathbf{\Psi}(r)=
     \begin{bmatrix}
         \psi_{c\overline{c}}(r)\\
         \psi_{M\overline{M}}^{(1)}(r)\\
         \vdots\\
         \psi_{M\overline{M}}^{(N)}(r)
     \end{bmatrix},
 \end{equation}
 where $\psi_{c\overline{c}}(r)$ is the wave function of $c\overline{c}$, $\Psi_{M\overline{M}}^{(i)}(r)$($1\le i\le N$) is the wave function  of the i-th meson-meson component. The normalization condition satisfied by the wave function is
 \begin{equation}
 \int\mathrm{d}\boldsymbol{r}\Psi^\dagger(r)\Psi(r)=\mathcal{P}(c\bar{c})+\mathcal{P}_1(M\bar{M})+\cdots+\mathcal{P}_n(M\bar{M})=1
 \end{equation}
 where we have defined the probility of finding the $c\bar{c}$ and $M\Bar{M}_i$ components in the state as
 \begin{equation}
    \mathcal{P}(c\overline{c})\equiv\int\mathrm{d}\boldsymbol{r}|\psi_{c\overline{c}}(r)|^2,
 \end{equation}
 and
 \begin{equation}
 \mathcal{P}_i(M\overline{M})\equiv\int\mathrm{d}\boldsymbol{r}|\psi_{M\overline{M}}^{(i)}(r)|^2.
 \end{equation}

\section{POTENTIAL MATRIX WITH SPIN-DEPENDENT TERMS}
In the last section, we have presented the main formalism for the diabatic approach. In this part, we shall give the specific form of the matrix elements of $V_{ij}(\boldsymbol{r},\boldsymbol{r}_0)$, which are needed to solve the Schrödinger equation of the system.
\subsection{$c\overline{c}$ potential}
 In this part, we shall discuss the interactions of charm quarks. For the element $V_{c\bar c}(r)$, as can be seen from Eq.(\ref{17}), it describes the static energy of the light field state corresponding to a pure $c\bar c$ state. Therefore, we can take the $V_{c\bar c}(r)$ as the conventional $c\overline{c}$ potential\cite{bali2001qcd}.

 There are many models for the effective potential of $c\overline{c}$, among which the non-relativistic potential model is relatively simple and widely used. The main part of this model contains a color Coulomb potential and a confinement potential \cite{eichten2008quarkonia}. In this paper, we will use the linear potential as the confinement potential. Hence the central potential can be written as
 \begin{equation}
    V_0^{(c\bar{c})}(r)=-\frac43\frac{\alpha_s}r+br,\label{20}
 \end{equation}
 where $\alpha_s$ and b are model parameters.

 In addition, we will further consider spin-dependent interactions in the potential in this work, which are not considered in Refs.\cite{PhysRevD.102.074002,Bruschini_2021}. Following the method in Ref.\cite{Lucha:1995zv}, we shall introduce three spin-dependent terms in Hamiltonian.

 Firstly, we consider the spin-spin contact hyperfine potential. This interaction is one of the spin-dependent terms predicted by one gluon exchange (OGE) potential. In this work, we take it as the Gaussian-smeared form
 \begin{equation}
    V_{SS}(r)=\frac{32\pi\alpha_s}{9m_c^2}\tilde{\delta}_{\sigma}(r)\vec{S}_{c}\cdot\vec{S}_{\bar{c}},\label{21}
 \end{equation}
 where $\vec{S}_{c}$ and $\vec{S}_{\bar{c}}$ are spin operators acting on the spin of quark and antiquark. We take $\tilde{\delta}_\sigma(r)~=~(\sigma/\sqrt{\pi})^3e^{-\sigma^2r^2}$ \cite{barnes2005higher}, and $\sigma$ and $m_c$ are model parameters.\par
 The remaining spin-dependent interactions are from the spin-orbit and tensor couplings between the $c\bar c$, which can also be deduced from the OGE potential. Using the leading-order perturbation theory, the spin-orbit potential $V_{SO}$ and the tensor potential are obtained as
 \begin{equation}
    V_{SO}(r)=\frac1{m_c^2}\left(\frac{2\alpha_s}{r^3}-\frac{b}{2r}{}\right)\vec{L}\cdot\vec{S}, \label{22}
 \end{equation}
 \begin{equation}
    V_{T}(r)=\frac1{m_c^2}\left(\frac{4\alpha_s}{r^3}\text{ T }\right),
\end{equation}
\begin{equation}
    \text{ T }=\frac{\vec{S}_{c}\cdot\vec{r}\vec{S}_{\bar{c}}\cdot\vec{r}}{r^2}-\frac{\vec{S}_{c}\cdot\vec{S}_{\bar{c}}}{3}
     \end{equation}
 where $\vec{L}$ is the orbit angular momentum operator, $\vec{S}$ is the total spin operator and T is the tensor operator.

 In the $|J,L,S\rangle$ basis, the elements of spin-orbit operator and spin-spin operator are diagonal. Furthermore, the off-diagonal elements of the tensor term are very small and can be neglected \cite{PhysRevD.29.110}. So the matrix elements involving spin operators can be given as \cite{barnes2005higher}
 \begin{equation}
     \langle\vec{\mathrm{S}}_c\cdot\vec{\mathrm{S}}_{\overline{c}}\rangle=\frac{1}{2}S^2-\frac{3}{4},
 \end{equation}
 \begin{equation}
     \langle\vec{\mathrm{L}}\cdot\vec{\mathrm{S}}\rangle=[J(J+1)-L(L+1)-S(S+1)]/2,
 \end{equation}
 \begin{equation}
     \langle^3\mathrm{L_J}|\mathrm{T}|^3\mathrm{L_J}\rangle =
     \begin{cases}
         -\frac{L}{6(2L+3)}, & J=L+1\\
         +\frac{1}{6},& J=L\\
         -\frac{L+1}{6(2L-1)},&J=L-1.
     \end{cases}
 \end{equation}\par
After including the spin-dependent terms, now the total interaction potential is
 \begin{equation}
     V_{00}(r)=V_{c\overline{c}}(r)=V_0^{(c\bar{c})}(r)+V_{SS}(r)+V_{SO}(r)+V_{T}(r).\label{27}
 \end{equation}
The values of the parameters ($m_c,\alpha_s,b,\sigma$) in the potential
 will be discussed in the next section. The adopted values of them are listed in Tab. \ref{tab:III}.

\subsection{Meson-meson potential}
 Next, we shall consider the interaction potential within meson-meson pair in Eq.\eqref{9}.
 As mentioned in Section II, one advantage of the diabatic approach is that with taking an appropriate value for the $r_0$ each state $|\zeta_i(\boldsymbol{r_0})\rangle$($1\le i\le N$) corresponds to a concrete pure meson-meson configuration. So the matrix elements of $V_{M\overline{M}}$ have clear physical meanings. Its diagonal elements  represent the internal interaction of each meson-meson pair, and the off-diagonal elements represent the couplings between different meson-meson pairs.

 For the off-diagonal elements of the $V_{ij}^{M\bar M}$, as previously mentioned, we will set them to be 0, i.e. neglecting the couplings between different meson-meson pairs. That is
 \begin{equation}
     V^{M\overline{M}}_{ij}(r)=0, ~i \ne j,1\leq i,j \leq N.
 \end{equation}\par
 For the diagonal elements, we use the potential as in Ref.\cite{PhysRevD.102.074002}, i.e.
 \begin{equation}
 V^{M\overline{M}}_{ii}(r)=T^{M\overline{M}}_i\equiv m_{M_1}^i+m_{\overline{M}_2}^i,~1\leq i\leq N,\label{29}
 \end{equation}
 where $m_{M_1}$ and $m_{\overline M_2}$ stand for the masses of the mesons in the meson pair. It means we also do not consider the interactions between mesons here.

\subsection{Mixing potential}
 Next, we shall discuss the mixing potential $V^{mix}(r)$ which couples $c\overline{c}$ with meson-meson pairs. The mixing potential used in this work is extracted from lattice quantum chromodynamics (LQCD) results\cite{Bulava_2019} following the approach in Ref.\cite{PhysRevD.102.074002}. Let's consider the coupling of $c\overline{c}$ and the $i$th meson pair. According to Ref.\cite{PhysRevD.102.074002}, there exists a crossing radius $r^c$, where the interaction potential of $c\overline{c}$ equals the threshold mass of the $i$th meson pair. So at $r=r_i^c$ we have
\begin{equation}
    V_{c\overline{c}}(r^c_{i})=T^{M\overline{M}}_i.
\end{equation}
 Furthermore, the mixing potential $V^{mix}$ satisfies the following conditions.
Firstly, when $r=r^c$ the mixing potential $V^{mix}(r)$ reaches its maximum value.    Secondly, when r is far from $r^c$, the mixing potential quickly approaches 0. Therefore, a reasonable choice for the $V^{mix}$ is a Gaussian function, and it can be taken as
 \begin{equation}
 V^{\mathrm{mix}}_i(r)=\frac\Delta2\mathrm{exp}\left\{-\frac{(V_{c\overline{c}}(r)-T^{M\bar{M}}_i)^2}{2(b\rho)^2}\right\},
 \end{equation}
 where $\Delta$ and $\rho$ are parameters representing the maximum value and the width of the Gaussian function, respectively. Here b is the same parameter as in Eq.\eqref{20}. The values for the parameters $\rho$ and $\Delta$ will be discussed and given below. Therefore, now we get the matrix elements of mixing potential in Eq.\eqref{14} as
 \begin{equation}
      V_{0i}(r)=V_{i0}(r)=V^{\mathrm{mix}}_i(r),~i\ne 0.\label{32}
 \end{equation}\par
As mentioned above, the mixing of $c\bar{c}$ and $M\bar{M}$ primarily occurs near the crossing radius $r^c$. In this region, the long-range part of $V_{c\overline{c}}(r)$(Eq.\ref{27}) will play a major role in the mixing potential $V^{\mathrm{mix}}_i(r)$. In our model, the spin-dependent terms are mainly short-range, and the mixing of different $c\bar{c}$ states due to the tensor terms is also neglected. Therefore, the inclusion of the spin interactions within the $c\bar{c}$ system shall not alter the form of the mixing potential $V^{\mathrm{mix}}_i(r)$ in Eq.(\ref{32}).
\section{RESULTS and discussions}
In this section, we present the numerical results and discussions. Firstly, in part A we shall discuss some ingredients needed to describe the charmonium states and the values of parameters in this work. Then, we will present the calculated results and discussions in part B.
\subsection{Description of charmonium(-like) states}
By using the method and formalisms discussed above, now we can study the  charmonium(-like) mesons with taking into account the coupling of $c\bar{c}$ with meson-meson pairs. Solving the Schrödinger equation by numerical methods, we can obtain the mass spectrum of the charmonium(-like) mesons. In Tab.\ref{tab:I}, we list the various meson-meson pairs considered in this work and their corresponding threshold masses.\par
 \begin{table}[h]
     \centering
      \caption{Low-lying open charm meson-meson channel considered in this work and their thresholds\cite{ParticleDataGroup:2024cfk}.}
     \begin{tabular}{cc}
          \hline\hline
          $M\Bar{M}$&$T^{M\Bar{M}}(MeV)$  \\
          \hline
          $D\bar{D}$& 3730\\
          $D\bar{D}^*(2007)$ & 3872\\
          $D_s^+D_s^-$&3937\\
          $D^*(2007)\bar{D}^*(2007)$&4014\\
          $D_s^+\bar{D}_s^{*-}$&4080\\
          ${D}_s^{*+}\bar{D}_s^{*-}$&4224\\
          \hline\hline
     \end{tabular}

     \label{tab:I}
 \end{table}
Heavy quark meson states are classified by their quantum numbers which include isospin($I$), G-parity($G$), total angular momentum($J$), parity($P$), and charge conjugation($C$). In this paper, we shall focus on the isoscalar meson states, i.e. $I=0$, which also means $G=C$.

Note that the considered diabatic potential is centrally symmetric, and there are no off-diagonal elements in the $|J,L,S\rangle$ basis for the spin-dependent operators (here $J,L,S$ represent the total angular momentum, orbital angular momentum, and total spin of the system). Therefore, for a system of $c\bar{c}$ or meson-meson states, as long as we know the $J, L, S$ of the system, we can deduce its $J^{PC}$ quantum numbers. The coupling of $c\bar{c}$ with meson-meson states can only happen when they share the same $J^{PC}$. In Tab.\ref{tab:II} we list the $J^{PC}$ quantum numbers and the corresponding orbital angular momentum of the $c\bar c$ and meson-meson states that can couple with each other.
\begin{table}[h]
    \centering
    \caption{Quantum numbers of the charmonium(-like) mesons considered in this work. The corresponding orbital angular momentum $l$ of various components are also given, and a blank space means that no possible orbital angular momentum $l$ exists.}
    \begin{tabular}{ccccc}
        \hline\hline
        $J^{PC}$ & $l_{c\overline{c}}$ & $l_{D_{(S)}\overline{D}_{(S)}}$ &$l_{D_{(S)}\overline{D}_{(S)}^*}$& $l_{D_{(S)}^*\overline{D}_{(S)}^*}$ \\
        \hline
         $1^{--}$& 0,2 &1 & 1& 1,3\\
         $2^{++}$& 1,3 &2 & 2& 0,2,4 \\
         $1^{++}$& 1 &  & 0,2& 2\\
         $0^{++}$& 1 & 0&  & 0,2\\
         $0^{-+}$& 0 & & 1& 1\\
         $1^{+-}$& 1 & & 0,2& 0,2\\
         \hline\hline
    \end{tabular}
    \label{tab:II}
\end{table}

 For a meson state with quantum numbers $J^{PC}$, by reading the possible configurations from Tab.\ref{tab:I}, we can construct the Schrödinger equation Eq.\eqref{33} for the system.
 The obtained coupled-channel Schrödinger equation describes the corresponding heavy meson system.

 Next, we need to discuss the values of the parameters used in this work. At present, we have six parameters in the potential matrix, i.e. $m_c$, $\alpha_s$, $b$, $\sigma$, $\rho$ and $\Delta$. Besides these six parameters, we still need one more new parameter $r_{cut}$ to cure the singularity problem due to the $\frac{1}{r^3}$ term in the potential as $r\to 0$. Here we follow the approach of Ref.\cite{PhysRevD.95.034026} with taking $\frac{1}{r^3}=\frac{1}{r_{cut}^3}$ in the region of $0<r<r_{cut}$ to resolve this problem. Therefore, we have seven parameters in total. For the $r_{cut}$ and the four parameters ($m_c,\alpha_s,b,\sigma$) associated with the $c\overline{c}$ interaction, we take their values from Ref. \cite{PhysRevD.95.034026}, where the same $c\bar c$ potential was adopted and the parameters were determined by fitting the masses of 12 well-established $c\bar c$ states.

 For the parameters $\rho$ and $\Delta$, which appear in the mixing potential, we basically follow the approach in Ref. \cite{PhysRevD.102.074002}. For the $\rho$, as argued in Ref.\cite{PhysRevD.102.074002}, the unquenched lattice QCD calculations rule out a large radial scale for the mixing and then the authors adopt $\rho=0.3$ fm in their calcultations. Here we follow their arguments and adopt the same value for the $\rho$. While, to account for the possible effects due to a different $c\bar c$ potential used in this work, we set $\Delta$ as a free parameter and obtain its value by fitting the mass of $\chi_{c1}(3872)$. All the values of the parameters are then determined and are collected in Tab. \ref{tab:III}.

 Now, we are ready to solve the coupled-channel Schrödinger equation of the system by numerical methods and obtain the spectrum of the heavy meson system. In this work, we use the renormalized Numerov algorithm to perform the numerical calculations, and the details of this method can be found in Ref. \cite{johnson1978renormalized}.

 \begin{table}[h]
     \centering
      \caption{Parameters adopted in the potential matrix.}
   \begin{tabular}{cc|cc}
        \hline\hline
         Parameters & Value & Parameters & Value  \\
         \hline
          $m_c$ (GeV) & 1.4830 &$r_{cut}$ (fm) & 0.202\\
          $\alpha_s$ & 0.5461 &$\rho$ (fm) & 0.3\\
          $b$ (Ge$\mathrm{V^2}$) & 0.1425 &$\Delta$ (GeV) & 0.116\\
          $\sigma$ (GeV) & 1.1384\\
          \hline\hline
     \end{tabular}

     \label{tab:III}
 \end{table}

\subsection{Spectrum and probabilities}
 By solving the coupled-channel Schrödinger equation (Eq.\eqref{33}), the mass spectrum  of the charmonium(-like) mesons and the probabilities of each component in them are obtained. The results are presented in Tab.\ref{tab:IV} and Tab.\ref{tab:V}, respectively.

 In Tab.\ref{tab:IV}, we show the calculated results of charmonium mass spectrum together with the corresponding experimentally observed states quoted from the Particle Data Group(PDG) book\cite{ParticleDataGroup:2024cfk}. Furthermore, the results of three other theoretical works are also presented for comparison. The results of Ref.\cite{PhysRevD.95.034026} are based on a   nonrelativistic quark model without considering the coupled-channel effects from the meson-meson pairs. The other two are taken from the works using the same diabatic approach but without considering the spin-dependent interactions between quarks\cite{PhysRevD.102.074002,Bruschini_2021}.

 From Tab.\ref{tab:IV}, one can see that our calculated masses can describe the experimental data quite well. Let's first discuss the lowest lying state of each $J^{PC}$ quantum numbers. It can be seen that our results are basically same as Ref.\cite{PhysRevD.95.034026}. This should not be surprising, since we use the same $c\bar c$ potential and the coupled-channel effects are very small due to their masses being far away from the open charm threshold. On the other hand, when compared to the results obtained in the diabatic approach\cite{PhysRevD.102.074002,Bruschini_2021}, we find that the results are improved significantly after including the spin-dependent interactions.

\begin{table}[h]
    \centering
     \caption{Calculated masses for charmonium(-like) states. The experimental results(Exp.) are taken from the PDG book\cite{ParticleDataGroup:2024cfk} in unit of MeV. The results from a conventional quark model\cite{PhysRevD.95.034026} and the results using diabatic approach in Refs.\cite{PhysRevD.102.074002,Bruschini_2021} are also presented for comparison.}
    \begin{tabular}{ccccccc}
        \hline\hline
         $J^{PC}$&name&Exp.& Ref.\cite{PhysRevD.95.034026}&Ref.\cite{PhysRevD.102.074002}& Ref.\cite{Bruschini_2021} &This work\\
         \hline
         $1^{--}$ &$J/\psi$ & 3096.9 & 3097&3082.4 & 3082.4&3097.2 \\
         & $\psi(2S)$ &  3686.1& 3679& 3664.2 &3658.8& 3669.9 \\
         & $\psi(3770)$ &  3778.1& 3787& 3790.2 & 3785.8 & 3782.8 \\
         & $\psi(4040)$ & 4039.6& 4078 & 4071 & & 4060.1 \\
         $2^{++}$& $\chi_{c2}(1P)$ &  3556.2& 3552& 3509.6 & 3508.7 &3550.4   \\
         & $\chi_{c2}(3930)$ &  3922.5& 3967& 3933.5& 3909.0&  3934.0  \\
         &  $\chi_{c2}(3P)$& &4310 & & 4006.6 &  4012.8\\
         $1^{++}$& $\chi_{c1}(1P)$ & 3510.7&3516 & 3510.0& 3509.8&  3515.0 \\
         & $\chi_{c1}(3872)$ & 3871.6 &3937 &3871.7 & 3871.5 & 3871.7 \\
         $0^{++}$& $\chi_{c0}(1P)$ & 3414.7&3415 &3509.1 & 3508.8 &  3416.8 \\
         & $\chi_{c0}(3860)$ &  3862& &  & &  3862.0  \\
         & $\chi_{c0}(3915)$ &  3922.1& 3869&3920.4 & 3918.9&    \\
         $0^{-+}$&$\eta_c(1S)$& 2984.1&2983 & & & 2984.6 \\
         & $\eta_c(2S)$ &3637.7 &3635& & & 3633.7 \\
         & $\eta_c(3S)$ & &4048 & & & 4039.4\\
         $1^{+-}$ &$h_c(1P)$ & 3525.4 & 3522& & & 3521.5\\
         & $h_c(2P)$ & & 3940& & & 3927.9\\
         \hline\hline
    \end{tabular}

    \label{tab:IV}
\end{table}

\begin{figure}
 \includegraphics{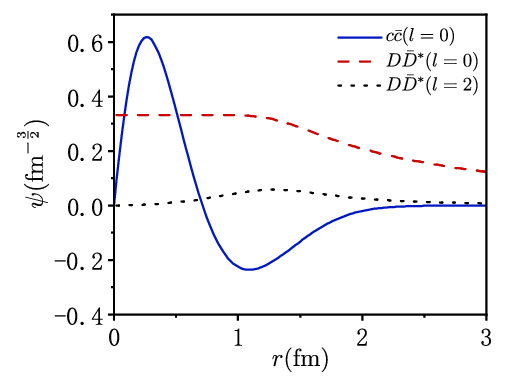}
 \caption{\label{fig1}Radial wave function of the calculated $0^+(1^{++})$ state with a mass of 3871.7 MeV.}
\end{figure}
 Next, let’s come to higher lying states. For the $1^{++}$ state $\chi_{c1}(3872)$, we need to note that, as mentioned above, we utilize this state to determine the parameter $\Delta$ in the mixing potential. From Tab.\ref{tab:V}, one can see that the probability of finding $D\overline{D}^*$ in this state is about $94\%$. Consequently, we can interpret this state as a predominantly molecular state. Its radial wave functions of various components are plotted in Fig.\ref{fig1}. The rms radius can be obtained as 6.55 fm, which is significantly large compared to its $c\overline{c}$ component with a rms radius of 1.15 fm.

\begin{figure}
 \includegraphics{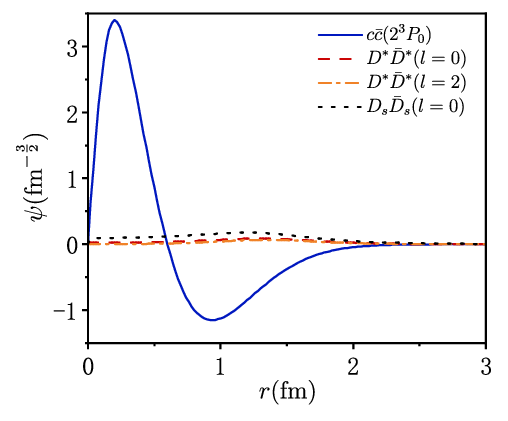}
 \caption{\label{fig2}Radial wave function of the calculated $0^+(0^{++})$ state with a mass of 3862.0 MeV.}
\end{figure}

 For the $0^{++}$ state $\chi_{c0}(3860)$\cite{ Belle:2017egg,Yu:2017bsj}, our calculated mass is 3862 MeV, which is exactly same as the central value suggested by PDG. Its radial wave functions of various components are drawn in Fig.\ref{fig2}. The resulting rms radius is 1.05 fm. The probabilities of various compenents in this state indicate that $\chi_{c0}(3860)$ is predominantly composed of $c\overline{c}$ (95$\%$).
 So it is not surprising that our calculated mass for $\chi_{c0}(3860)$ is similar to the quark model result\cite{PhysRevD.95.034026}. This also means that the conclusion on $\chi_{c0}(2P)$ state still depends on the employed quark model in our work.
According to PDG\cite{ParticleDataGroup:2024cfk}, there are two mesons appearing at around 3900 MeV, i.e. $\chi_{c0}(3860)$ and $\chi_{c0}(3915)$. While, according to Refs.\cite{ortega2018charmonium,PhysRevLett.115.022001,Guo:2012tv}, it remains unclear whether $\chi_{c0}(3915)$ should be assigned to the $0^{++}$ or $2^{++}$ state. Our results suggest that the $\chi_{c0}(3860)$ can take the place of the $\chi_{c0}(2P)$ state. At the same time, since we do not get another state with $J^{PC}=0^{++}$ having mass around 3915 MeV, it may be looked as the support for the argument that the $\chi_{c0}(3915)$ is not a $0^{++}$ state. On the other hand, it is worth mentioning that recent LHCb  results\cite{LHCb:2020bls} indicate the simultaneous existence of the $\chi_{c0}(3930)$ and $\chi_{c2}(3930)$ states. Therefore, a thorough understanding the $\chi_{c0}$ states around the 3900 MeV still needs further experimental and theoretical efforts.

\begin{figure}[h]
 \includegraphics{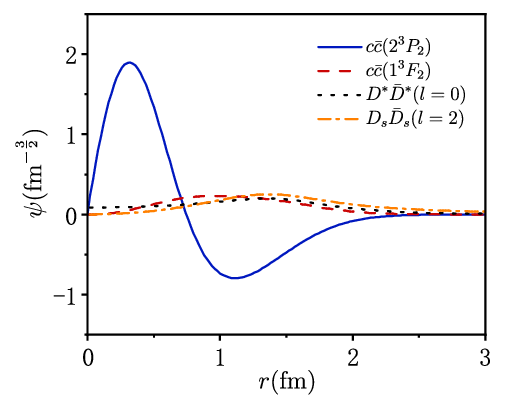}
 \caption{\label{fig3}Radial wave function of the calculated $0^+(2^{++})$ state with a mass of 3934.0
MeV.}
\end{figure}

 For the first excited $2^{++}$ state, as indicated in Tab.\ref{tab:V}, both $D^*\overline{D}^*$ and $D_s\overline{D}_s$ contribute approximately $10\%$ in this state. Experimentally, the mass of $\chi_{c2}(3930)$ is $3922.5\pm 1.0$ MeV. In comparison to the pure $c\bar c$ results, the inclusion of the coupled-channel effects significantly improve the results. The radial wave functions of its components are drawn in Fig.\ref{fig3}, and a rms radius of 1.35 fm can be obtained.

\begin{figure}[h]
 \includegraphics{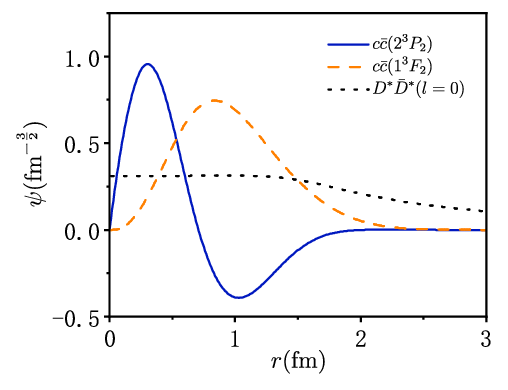}
 \caption{\label{fig5}Radial wave function of the calculated $0^+(2^{++})$ state with a mass of 4012.8
MeV.}
\end{figure}

 Regarding the second excited state with a mass of 4012.8 MeV of the $2^{++}$ spectrum, there is no corresponding meson state has been observed experimentally. In our model, this is a typical unconventional state. It is composed of the $D^*\bar D^*$(51$\%$) and $c\overline{c}$(48$\%$) components. Its radial wave functions are shown in the Fig.\ref{fig5}. The corresponding rms radius is 2.33 fm. Compared to the results obtained using the quark model in \cite{PhysRevD.95.034026}, the threshold effects of the $D^*\bar D^*$ channel reduce its mass by about 300 MeV. Consequently, the experimental search of this state can offer further verification of the approach used in this work.

 For the states of the $1^{--}$ spectrum, our results are in general closer to the experimental values than those obtained without considering spin-dependent interactions. For the first excited state, the $c\overline{c}$ component plays a dominant role ($96\%$), while the role of the $D\overline{D}$ component is minor (4$\%$). The resulting mass is 3669.9 MeV, which is smaller than the result of the conventional $c\bar c$ model, and the experimental value for this state($\psi(2S)$) is 3686.097 $\pm$0.010 MeV. For the second excited state, the $c\overline{c}$ component remains dominant and is almost a pure D-wave state(98$\%$), which is similar to the results of the quark model\cite{PhysRevD.95.034026}. The computed mass for this state is 3782.8 MeV, which is in good agreement with the experimental value for the meson $\psi(3770)$.

\begin{figure}[h]
 \includegraphics{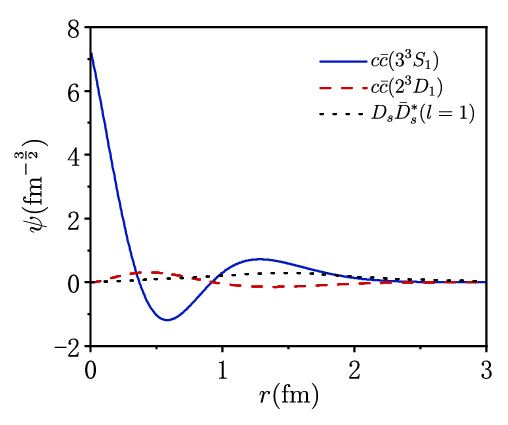}
 \caption{\label{fig6}Radial wave function of the calculated $0^-(1^{--})$ state with a mass of 4060.1
MeV.}
\end{figure}

 For the third excited state in the $1^{--}$ spectrum, the calculated mass is 4060.1 MeV and the contribution of the $D_s\overline{D}^*_s$ component in this state is significant(20$\%$). Its mass is close to the experimental value for the meson $\psi(4040)$ with its mass being $4039.6\pm 4.3$ MeV. The radial wave functions of this state are plotted in Fig.\ref{fig6}, and its rms radius is obtained as 1.42 fm.

 \begin{table*}
     \caption{Masses and probabilities of various components for the  charmonium(-like) states with different $J^{PC}$. The $c\overline{c}$ probabilities from different values of $l_{c\bar c}$(see Tab.II) are presented in parentheses. A missing entry under a meson-meson configuration means that the corresponding component gives negligible (probability inferior to 1$\%$) or no contribution at all to the state.}
    \begin{tabular}{llllllllll}
        \hline\hline
       $J^{PC}$&Mass(MeV) & $c\overline{c}$ &$D\overline{D}$ &$D\overline{D}^*$&$D_s\overline{D}_s$
       & $D^*\overline{D}^*$ & $D_s\overline{D}^*_s$ & $D^*_s\overline{D}^*_s$\\
         \hline
         $1^{--}$  &  3097.2 & (100 , 0)$\%$ \\
          & 3669.9 & (95 , 1)$\%$ & 4$\%$ &1$\%$ &  &   &  & \\
          &  3782.8 & (0 , 98)$\%$& & 2$\%$ &  &   &  &  &\\
          &  4060.1 &(74 , 5)$\%$& & & & & 20$\%$ & 2$\%$\\
         $2^{++}$& 3550.4& (100 , 0)$\%$ & & & &  \\
          & 3934.0 & (69 , 6)$\%$ & & & 13$\%$ &11$\%$ & 1$\%$ & \\
          &  4012.8 & (12 , 36)$\%$ & & & &51$\%$& & \\
         $1^{++}$& 3515.0 & 100$\%$ & &\\
          &  3871.7 &6$\%$ & & 94$\%$& \\
         $0^{++}$&  3416.8  &100$\%$& & \\
          &  3862.0 & 95$\%$& & & 4$\%$& 1$\%$&\\
         $0^{-+}$ & 2984.6&100$\%$& & \\
          &3633.7 &100$\%$& & \\
          &4039.4 &92$\%$& & & & &7$\%$&1$\%$\\
         $1^{+-}$ & 3521.5 &100$\%$& & \\
         &3927.9 & 93$\%$ & & & & 4$\%$ &2$\%$& \\
          \hline\hline

    \end{tabular}

    \label{tab:V}
\end{table*}

 For the low lying $\eta_c(nS)(0^{-+})$ states with $n=1,2$ and the state $h_c(1P)(1^{+-})$, they are well described by the conventional quark model. Our results are basically consistent with the quark model  \cite{PhysRevD.95.034026} and experimental results.
 While, similar to Ref.\cite{PhysRevD.95.034026}, we have also predicted two new states, whose masses are 4039.4 MeV with $J^{PC}=0^{-+}$ and 3927.9 MeV with $J^{PC}=1^{+-}$. The meson-meson threshold effect have a small but non-negligible impact ($<10\%$) on the results, which make the predicted masses of these two mesons a little smaller than the predictions of the conventional quark model\cite{PhysRevD.95.034026}.  For these states, further experimental information are still needed to verify the predictions.

As a final part, we present some discussions on the approximations adopted in this work. It should be noted that following Ref.\cite{PhysRevD.102.074002} our model can only treat bound state problem and at the energies above certain meson pair thresholds the mixing with these meson pairs are simply ignored. Due to this simplification, in the present model we can not study the coupling or mixing of $c\bar c$ state with the meson pair at energies above their mass thresholds. We remind the readers that possible states originated from such dynamics are not considered in this work.

Furthermore, it is also worth mentioning that in the present work we use the same coupling strength $\Delta$ for both
the charmed meson and charmed-strange meson channels in the mixing potential. To estimate the uncertainties due to this approximation, we have tried to further consider a factor of $\frac{m_{q}}{m_s}$ with $m_q=0.33$ GeV and $m_s=0.5$ GeV, which is used in the $^3P_0$ model to account for the difference of the couplings with different flavors\cite{Kalashnikova:2005ui}, for the coupling strength $\Delta$ of the charmed-strange meson channels. The calculated results show that including this factor may cause significant effects on the states lying very close to the meson pair threshold. For example, the $2^{++}$ state with a mass of 3934 MeV is very sensitive to this change, because
it lies very close to the $D_s^+D_s^-$ threshold. In fact, when using the new coupling strength, this bound state does not exist any more. While for other states, this change only causes moderate effects on the results.
 \section{summary}
In summary, by employing the diabatic approach and solving the coupled-channel Schrödinger equations with taking into account spin-dependent interactions in the $c\bar c$ system, we obtain the masses of charmonium spectrum below 4.1 GeV. Furthermore, we also study the probabilities of finding various components, i.e. $c\bar c$ or meson-meson pairs, in those states. Compared to previous studies utilizing the diabatic approach\cite{PhysRevD.102.074002,Bruschini_2021}, we find our results can describe the charmonium spectrum better after including spin-dependent interactions. In our results, the $\chi_{c0}(3860)$ and $\psi(3770)$ can be looked as the candidates for the $\chi_{c0}(2P)$ and $\psi(1D)$ states, respectively. In addition, our calculations show that the $\psi(4040)$ and $\chi_{c2}(3930)$ may have significant molecular components. It is also interesting to note that in our work we only get the $\chi_{c0}(3860)$ state, which differs from the $\chi_{c0}(3915)$ reported in Refs.\cite{PhysRevD.102.074002,Bruschini_2021}. Since there are still disputes on the nature of the $\chi_{c0}(3915)$\cite{ortega2018charmonium,PhysRevLett.115.022001,Guo:2012tv,LHCb:2020bls}, further experimental and theoretical efforts are still needed to understand the nature of the $\chi_{c0}(3915)$.

\begin{acknowledgements}
We acknowledge the support from the National Natural Science Foundation of China under Grant No.U1832160 and the Natural Science Foundation of Shaanxi Province under Grant No.2024JC-YBMS-010.
\end{acknowledgements}


\end{document}